\documentclass[onecolumn,showpacs,superscriptaddress,floatfix,preprintnumbers,amssymb,amsmath]{revtex4}

\usepackage{graphicx}
\usepackage{dcolumn}
\usepackage{bm}
\usepackage[latin1]{inputenc}
\usepackage[mathscr]{eucal}
\usepackage{epsfig}
\usepackage{rotating}

\begin{document}

\title{Quantized current of a hybrid single-electron transistor with superconducting leads and a normal-metal 
island}

\author{Antti Kemppinen}
\affiliation{Centre for Metrology and Accreditation (MIKES), P.O. Box 9, FIN-02151 Espoo, Finland}
\author{Matthias Meschke}
\affiliation{Low Temperature Laboratory, Helsinki University of Technology, P.O. Box 3500, FIN-02015 TKK, Finland}
\author{Mikko M\"{o}tt\"{o}nen}
\affiliation{Low Temperature Laboratory, Helsinki University of Technology, P.O. Box 3500, FIN-02015 TKK, Finland}
\affiliation{Department of Applied Physics/COMP, Helsinki University of Technology, P.O. Box 5100, FIN-02015 TKK, Finland}
\author{Dmitri V. Averin}
\affiliation{Department of Physics and Astronomy, Stony Brook University, SUNY, Stony Brook, NY 11794-3800, USA}
\author{Jukka P. Pekola}
\affiliation{Low Temperature Laboratory, Helsinki University of Technology, P.O. Box 3500, FIN-02015 TKK, Finland}

\begin{abstract}
We discuss the operation of the
superconductor--insulator--normal-me\-tal--insulator--superconductor (SINIS) turnstile. This
voltage-biased hybrid
single-electron transistor (SET) provides current quantization even with only one
radio-frequency (rf) control parameter, namely the gate voltage of the single island.
We give an overview of the main error mechanisms of the turnstile and consider its feasibility as a quantum 
current standard. We also present experimental results of pumping with the SINIS structure which show decreased 
leakage current compared to earlier measurements with the opposite NISIN structure.
\end{abstract} 


\maketitle

\section{Introduction}
\label{intro}

A device that can transfer electrons one by one or more generally $n$ electrons at frequency $f$ could 
be used as a quantum current standard with magnitude $I=nef$. The first proposals and devices of this kind 
were based on normal-metal tunnel junctions~\cite{Averin1986,Geerligs1990,Pothier1992}.
The most precise electron pump
built todate reached the relative accuracy $10^{-8}$. It consisted of a SET structure with seven
metallic tunnel junctions and six rf-controlled islands between them~\cite{Keller1996}. It
was used as a quantum capacitance standard~\cite{Keller1999}, but the $RC$ time constants of 
the tunnel junctions limited the maximum current on picoampere level which is not enough for a 
practical current standard. Specifically, at least 100~pA is required for closing the so-called quantum 
metrological triangle. This experiment compares the electrical quantum standards of voltage, 
resistance and current against each other via Ohm's law. It yields a consistency check for the 
fundamental constants $e$ and $h$~\cite{Likharev1985}. The Josephson voltage standard and the quantum-Hall 
resistance standard are routinely used in metrology institutes worldwide, but the current standard is the 
missing link. Devices generating higher currents than the SET pump, based on superconducting
devices~\cite{Niskanen2003,Vartiainen2007,Mooij2006}, surface acoustic waves~\cite{Shilton1996} and 
semiconducting quantum dots~\cite{Blumenthal2007,Kaestner2007,Kaestner2008,Fujiwara2008} have been proposed and 
realized. However, none of these devices has reached metrological accuracy.

\begin{figure}
\resizebox{0.4\columnwidth}{!}{
  \includegraphics{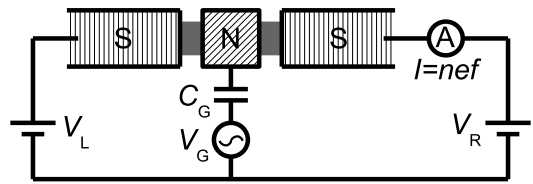} }
\caption{Schematic picture of the SINIS turnstile. A constant voltage is applied between the source and the 
drain. Pumping occurs with a single rf gate voltage control. Turnstile operation is possible also with the 
opposite hybrid structure with normal-metal leads and a superconducting island.}
\label{fig:pumppukaavio}
\end{figure}

In 2007, a new concept based on a hybrid SET structure was proposed and demonstrated in
experiments~\cite{Pekola2008}. The 
hybrid SET consists of either superconducting (S) leads connected via insulating (I) tunnel barriers
to a normal-metal (N) island or vice versa, see Fig.~\ref{fig:pumppukaavio}. When a constant bias voltage is 
applied over the SINIS/NISIN structure (e.g.~$V_\mathrm{L}=V/2$, $V_\mathrm{R}=-V/2$)
and the electrical potential of the island is controlled with an rf gate voltage signal, the 
structure acts as a turnstile. In the first
experiments~\cite{Pekola2008}, a NISIN turnstile showed robust current quantization. Theoretically, however,
the SINIS version promises higher accuracy. Detailed analysis predicts, that metrological accuracy 
could be reached with the simple SINIS structure with only one rf control parameter and two tunnel
junctions~\cite{Averin2008}. In this paper, we illustrate the principle of operation of the SINIS turnstile.
We present an overview of the error mechanisms and discuss the experimental challenges they pose. Finally,
we show experimental results of the SINIS turnstile with improved leakage properties.

\section{Theory of tunneling in Coulomb blockaded SINIS devices}
\label{sec:theory}
\begin{figure}
\resizebox{0.8\columnwidth}{!}{
  \includegraphics{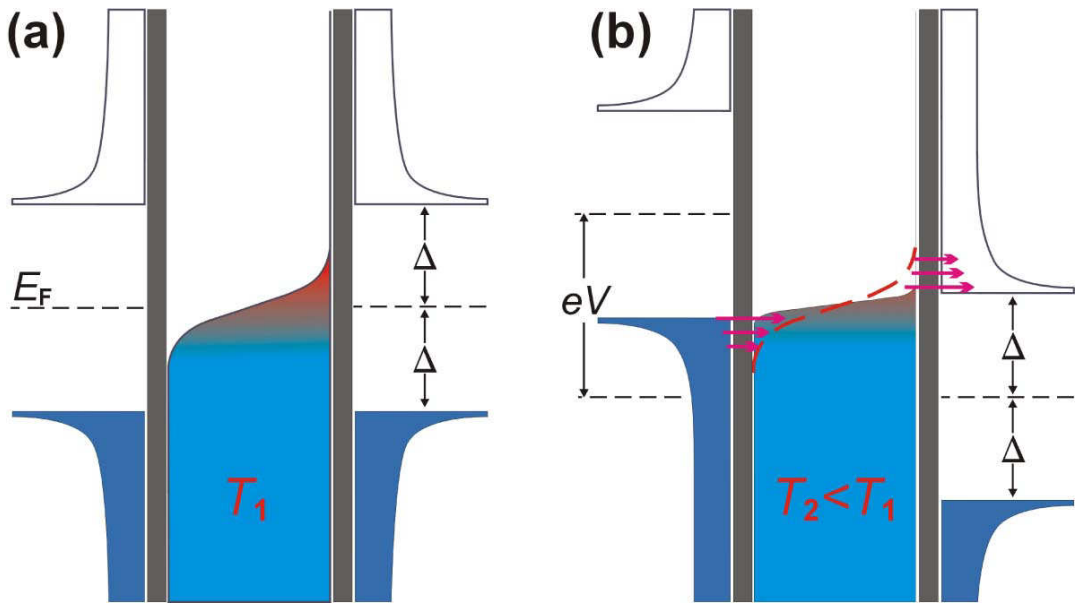} }
\caption{Schematic band diagram of the SINIS structure biased with (a) zero voltage and (b) voltage $V$ 
across the structure. For the superconducting leads, we show the BCS density of states which is zero in the 
range $E_\mathrm{F}\pm\Delta$ where $E_\mathrm{F}$ is the Fermi level and $\Delta$ is the BCS energy gap.
Close to the edges of the gap, the density of states 
approaches infinity. The states below the gap are almost perfectly occupied, and those above the gap are 
empty. In the normal metal, the Fermi function describing the occupancy of the states is shown instead of 
the density of states, which is roughly constant within the narrow energy range presented here. For these illustrations,
the gap has been narrowed for clarity. Applying voltage across 
the sample changes the Fermi levels of the superconductors by $\pm eV/2$. In (b), $eV$ is slightly below
$2\Delta$, and the normal metal cools down.}
\label{fig:bands}
\end{figure}

The tunneling rates of the SINIS turnstile are determined by the effects of the Coulomb blockade and the 
electronic structures of the superconductor leads and the normal-metal island. The latter effect can be 
understood qualitatively from the energy band diagram presented 
in Fig.~\ref{fig:bands}. Voltage $V$ over the structure shifts the Fermi levels of the superconductors
by $\pm eV/2$. At low temperatures, the Fermi distribution is almost like a step function. Tunneling can occur 
between states with same energy. Hence, at low voltages, electrons cannot tunnel from the occupied states 
of the normal metal to the superconductors, where the corresponding energy levels are either forbidden or
occupied. Similarly, there are no free states in the normal metal at occupied energy range of the 
superconductors. Therefore, the BCS gap $\Delta$ causes a voltage range $-2\Delta<eV< 2\Delta$ where the current 
through the device is very small.

In Fig.~\ref{fig:bands}(b), the structure is biased close to the edge of the gap. Electrons can tunnel from the
left superconductor to the states slightly below the Fermi energy of the normal metal filling cold holes, 
whereas hot electrons above the Fermi level of the normal metal can tunnel to the quasiparticle states of the 
right superconductor. The electron-electron interactions maintain the Fermi distribution, but it narrows,
which means that the normal metal cools down. Indeed, SINIS structures are routinely used as
microcoolers~\cite{Giazotto2006}. The Coulomb-blockaded structure used in the SINIS turnstile has also
been proposed for a radio-frequency single-electron refrigerator \cite{Pekola2007} and has been used as
a heat transistor~\cite{Saira2007}. The cooling effect may also improve the accuracy of the
turnstile.

The total capacitance of the island is $C_\Sigma=C_\mathrm{L}+C_\mathrm{R}+C_\mathrm{g}+C_0$, where
$\mathrm{L}, \mathrm{R}$, and $\mathrm{g}$ refer to the left and right tunnel junctions and to the gate,
respectively. The self-capacitance of the island is $C_0$. We normalize all energies by the BCS gap. The
normalized charging energy of the island is
$\epsilon_\mathrm{ch}=\epsilon_\mathrm{C}(n+n_\mathrm{g})^2$, where $n$ is the number of extra electrons
on the island and $n_\mathrm{g}=C_\mathrm{g}V_\mathrm{g}/e$ is the amount of charge in units of $e$
induced on the island by the gate. The unit of charging energy is
$\epsilon_\mathrm{C}=e^2/2C_\Sigma\Delta$. The change in energy for an electron tunneling to ($+$) or from ($-$) 
the island through junction $i$ is
\begin{equation}\label{ech}
\epsilon_n^{i,\pm}=\pm 2\epsilon_\mathrm{C}(n+n_\mathrm{g}\pm 1/2)\pm (v_i-\theta ).
\end{equation}
Here, $v_i=eV_i/\Delta$ is the normalized bias voltage of the junction (see Fig.~\ref{fig:pumppukaavio}) and
$\theta=(C_\mathrm{L}v_\mathrm{L}+C_\mathrm{R}v_\mathrm{R})/C_\Sigma$ is the offset to the island 
potential from the bias voltages.

Within the orthodox theory of sequential tunneling \cite{Averin1986}, the tunneling rates to and from the 
island are proportional to the numbers of empty and full states in the leads and on the island:
\begin{equation}\label{rates} \begin{array}{l}
\Gamma^{i,+}=\frac{\Delta}{e^2R_{\mathrm{T},i}}\int d\epsilon n_\mathrm{S}(\epsilon)f_\mathrm{S}(\epsilon)[1-f_\mathrm{N}(\epsilon-\epsilon^{i,+})]\\
\Gamma^{i,-}=\frac{\Delta}{e^2R_{\mathrm{T},i}}\int d\epsilon n_\mathrm{S}(\epsilon)f_\mathrm{N}(\epsilon+\epsilon^{i,-})[1-f_\mathrm{S}(\epsilon)]. \end{array}
\end{equation}
Here, $R_{\mathrm{T},i}$ is the tunneling resistance of the junction, and $f_\mathrm{S}$ and $f_\mathrm{N}$ are 
the Fermi distributions of the leads and the island, respectively. The distribution functions are identical, 
but the island can be at different temperature than the leads. The density of states in the
superconductors is ideally $n_\mathrm{S}(\epsilon)=|\epsilon|/\sqrt{\epsilon^2-1}$.

In the steady state, the probability of finding $n$ extra electrons on the island, $P(n)$, is constant, and the
net probability of transition between adjacent states is zero. Hence, we can solve the steady state from
the master equation
\begin{equation}\label{master}
[\Gamma^{\mathrm{L},+}(n)+\Gamma^{\mathrm{R},+}(n)]P(n)
=[\Gamma^{\mathrm{L},-}(n+1)+\Gamma^{\mathrm{R},-}(n+1)]P(n+1).
\end{equation}
The current through the junction L is simply
\begin{equation}\label{virta1}
I_\mathrm{L}=-e\sum_{n=-\infty}^{\infty}P(n)[\Gamma^{\mathrm{L},+}(n)-\Gamma^{\mathrm{L},-}(n)].
\end{equation}
In the steady state $I_\mathrm{L}=I_\mathrm{R}$.

\begin{figure}
\resizebox{0.9\columnwidth}{!}{
  \includegraphics{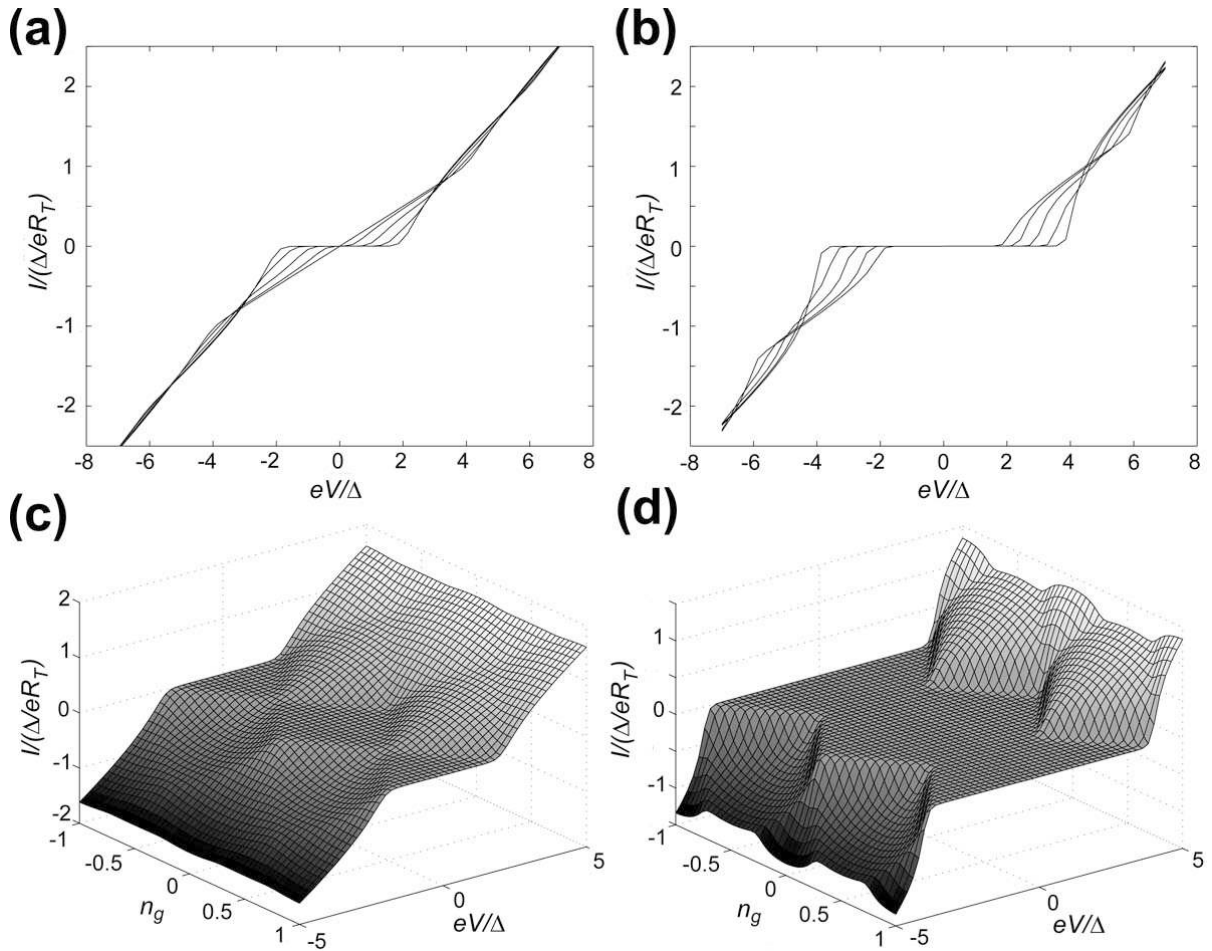} }
\caption{Theoretical IV curves of a normal (a) and a hybrid (b) SET with $n_\mathrm{g}$ between 0
(closed) and 0.5 (open). Here, $E_\mathrm{C}=\Delta$. The SINIS SET has a high zero-bias resistance even in
the gate open state due 
to the BCS gap. The graph (c) shows the current of the normal SET as a function of bias voltage and 
gate charge. The diamond-like stability regions with negligible current correspond to charge states -1, 
0 and 1. In the hybrid case (d), the stable regions are expanded by the BCS gap and they overlap.}
\label{fig:teorIV}
\end{figure}

Theoretical current-voltage (IV) curves calculated with these equations are presented in
Figs.~\ref{fig:teorIV}(a) and \ref{fig:teorIV}(b) for the NININ (normal SET) and SINIS (hybrid SET) cases, 
respectively. The main difference 
is that the BCS gap enlarges the areas where the charge states are stable and hence the current is ideally zero.
On this scale, the IV curves of the SINIS and NISIN structures are the same. The case of a superconducting
(SISIS) SET is not shown here, but it has the significant difference to the presented curves that the 
supercurrent can pass the structure at zero voltage.

In the literature, the density of states in a superconductor is often written as
\begin{equation}\label{nS}
n_\mathrm{S}(\epsilon)=\left| \mathrm{Re}\frac{\epsilon+i\gamma}{\sqrt{(\epsilon+i\gamma)^2-1}} \right|.
\end{equation}
The physical meaning of the $\gamma$ parameter is the Cooper-pair breaking rate of the
superconductor~\cite{Dynes1984}. In the absence of Coulomb blockade, pair breaking yields a finite linear
conductance $G_0=\gamma /R_\mathrm{T}$, and hence a leakage current in the regime $|eV/\Delta|< 2$. Ideally, the 
sub-gap conductance would be zero. In the Coulomb-blockaded case with gate open, $G_0=\gamma /2R_\mathrm{T}$. 
However, there are also other processes that lead to finite conductance, like the Andreev
reflection~\cite{Averin2008}. The details of the leakage processes have usually been neglected,
and $\gamma$ has been used as a general phenomenological leakage parameter like in Ref.~\cite{Pekola2008}.
A typical amount of leakage is $\gamma= 10^{-4}\ldots 10^{-3}$. Since the extremely low leakage current within 
the BCS gap is essential for the accuracy of the turnstile, we will discuss the leakage processes in
Sec.~\ref{sec:error}.

\section{Principle of operation}
\label{sec:principop}

\begin{figure}
\resizebox{1\columnwidth}{!}{
  \includegraphics{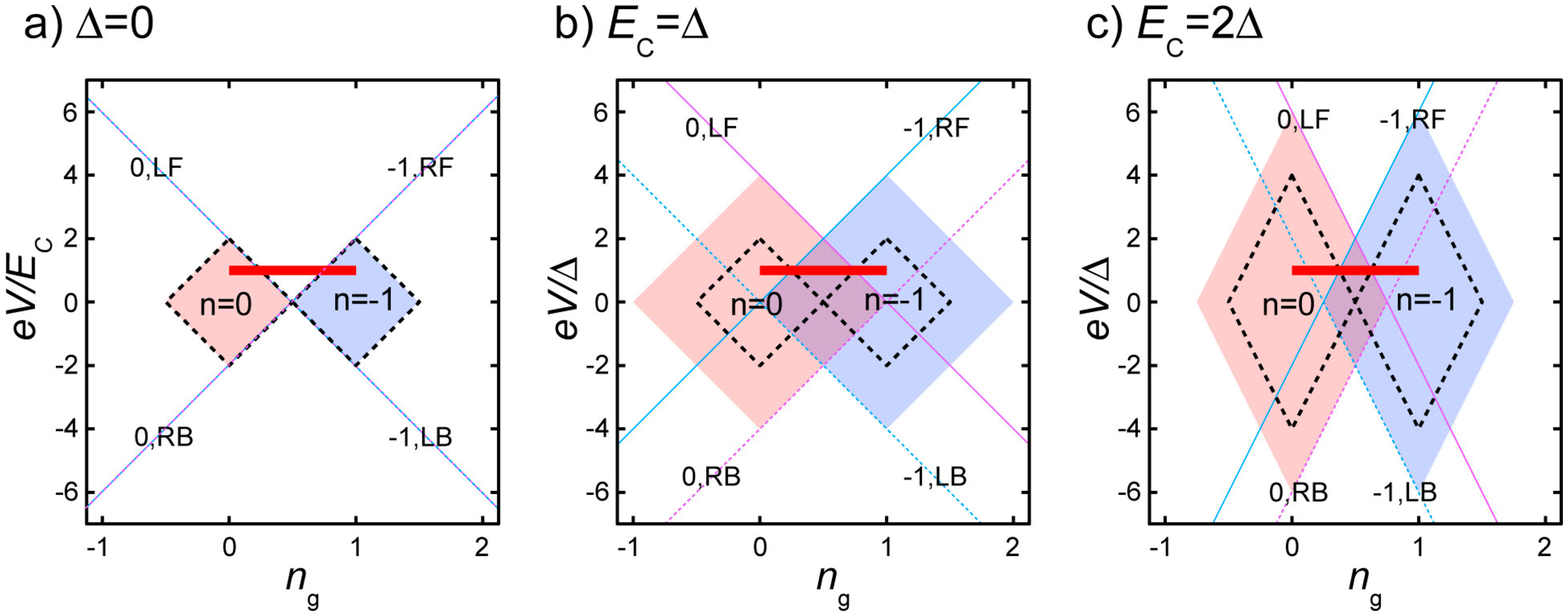} }
\caption{Schematic picture of pumping (a) with the normal SET, (b) with the hybrid SET with equal 
charging and BCS gap energies, and (c) with the hybrid SET with $E_\mathrm{C}=2\Delta$. The shaded
diamond-like areas are the stability regions of the charge states $n=0$ and $n=-1$. The edges of the
normal SET stability regions are drawn to all figures with dashed black lines. The thin lines
represent the transition thresholds from states $n=0$ and $n=-1$ by tunneling through the
left (L) or the right (R) junction in the wanted forward (F, solid line) or unwanted backward (B, 
dashed line) direction. We define the bias voltage to be positive in the left electrode. Hence, the
current flows from left to right, and the forward tunneling direction is from right to left.
The thick line corresponds to pumping with constant bias voltage
$eV/\Delta=1$ and the gate charge varying between the values $n_\mathrm{g}=0$ and $n_\mathrm{g}=1$.}
\label{fig:timantit}
\end{figure}

The operation of the SINIS turnstile can be explained with the help of
Fig.~\ref{fig:timantit}. In the normal case \ref{fig:timantit}(a), the
stability regions arising from the Coulomb interaction barely touch each other. The 
pumping signal illustrated with the thick line passes the zone outside the stability regions, where current can 
flow through the device. Without the bias voltage, there would be no preferred direction of 
tunneling, e.g.~transition from the state $n=0$ to $n=-1$ can occur by tunneling through the left 
junction in the forward direction or through the right junction in the backward direction. Hence, the
normal SET cannot act as a turnstile even in principle.

In the hybrid case shown in Figs.~\ref{fig:timantit}(b) and \ref{fig:timantit}(c), the stability regions are 
expanded by 2 units on the $eV/\Delta$ axis and by $\Delta/2E_\mathrm{C}$ on the $n_\mathrm{g}$ axis. Hence, the 
regions overlap. In the overlapping region, the system stays in the initial stable state. Now, let us consider 
the situation, where the device is 
biased to $eV/\Delta=1$, and we start to increase $n_\mathrm{g}$ from zero to unity. The device stays in the 
region, where at least one of the charge states is stable. The transition from the state 0 to $-1$ could happen 
due to an electron tunneling forwards through the left junction or backwards through the right junction.
Because of the bias, we meet the threshold for forward tunneling well before that of backward tunneling.
When the gate charge is again reduced to zero, the bias voltage causes a preferred direction of 
tunneling in a similar fashion.

One can also extend the gate voltage signal to span over several charge states. If, e.g., the path in
Fig.~\ref{fig:timantit} is extended to $n_\mathrm{g}=2$, the system would transfer from the state 0 to $-2$ by 
two sequential forward tunneling events through the left junction, and then back to the state 0 by two forward 
events through the right junction. In this process, two electrons are transferred per cycle. Current 
plateaus up to $I=10ef$ were obtained already in the first experiments~\cite{Pekola2008}.

One of the key benefits of the hybrid design is that it allows current quantization with a single island.
A pump with $N$ islands has $N$ unique fluctuating background charges that must be compensated by applying
offset voltages to the gate signals. Employing many rf controls to a dilution refrigerator is also a 
considerable 
engineering effort. In the hybrid case, this effort can be used to connect $N$ turnstiles in parallel which 
increases the current and decreases the relative statistical error. Also the maximum operation frequency,
limited by the $RC$ time constants, is lowered by the number of junctions. Therefore, the hybrid structures can 
be used to create roughly $N^2$ times the current of an $N$-island pump with the same amount of complexity.

Also the normal-state turnstile of Ref.~\cite{Geerligs1990} has a dc bias and only one rf gate control.
However, it requires at least two junctions on each side of the gate-controlled island. It means that there are 
at least two extra islands whose charge is either not well defined or they require extra controls. Four 
junctions cannot be biased optimally with the two control parameters~\cite{Jensen1992}. In the normal-state
turnstile, dc bias always leads to self-heating unlike in the SINIS case (see Fig.~\ref{fig:bands}(b)).

\section{Error sources}
\label{sec:error}

The error mechanisms of the SINIS turnstile have been analyzed theoretically in detail in
Ref.~\cite{Averin2008}. Here, we give a brief overview of them. The error sources can be divided into three 
categories: errors related to the pumping speed, thermally activated errors in tunneling, and several kinds of 
leakage processes.

The turnstile can be operated with any gate waveform oscillating around $n_\mathrm{g}=1/2$ with amplitude large 
enough to pass the tunneling thresholds. An obvious choice is the sine wave, which was used in the first 
experiments~\cite{Pekola2008}, and which is experimentally the easiest to accomplish. However, a square-wave 
signal would be optimal, because the time spent between the optimal gate charge values increases the 
probability of unwanted transitions. Also, the time spent at the extreme values decreases the probability of 
missed tunneling, which is $\exp (-\Gamma^{i,\pm}/2f)$ for the square-wave signal. Here, $\Gamma^{i,\pm}$ is 
the tunneling rate of Eq.~(\ref{rates}) of the wanted transition. At zero temperature,
$\Gamma^{i,\pm}=(\Delta/R_{\mathrm{T},i}e^2)\sqrt{(\epsilon_n^{i,\pm})^2-1}$, where the electrostatic energy change 
$\epsilon_n^{i,\pm}$ of Eq.~(\ref{ech}) is roughly proportional to $1/C_\Sigma$. Hence, the probability of 
missed tunneling depends on the $RC$ time constant as in normal-state devices. If the transition 
speed of the square-wave signal approaches $\Delta/h\approx 50$~GHz, the transistor can be excited to higher 
energy states, which causes errors. Realistic pumping frequencies are, however, not close to this limit.

The probability of an electron tunneling through the wrong junction due to thermal activation depends on the
energy difference to the zero-bias case with no preferred tunneling direction: $e^{-eV/k_\mathrm{B}T}$. This 
error leads to no net charge transferred during the cycle. On the other hand, if the bias point is 
close to the edge of the stability regions, $eV/\Delta=2$, an excitation can cause an extra electron
to transfer during the cycle. Likewise, this error probability depends on the energy difference:
$e^{-(2\Delta-eV)/k_\mathrm{B}T}$. Combining these equations, we get $eV/\Delta=1$ as the optimum bias voltage, 
and a thermal error probability $e^{-\Delta/k_\mathrm{B}T}$. The combined thermal error probability is less than
$10^{-8}$ at realistic temperatures of about 100~mK and with the gap of aluminum, $\Delta/k_\mathrm{B}=$2.3~K.

The most severe error source of the turnstile appears to be the different leakage processes. They involve
elastic and inelastic higher-order tunneling processes. In the NISIN case, the dominant process is the elastic 
electron cotunneling that limits the accuracy to about $10^{-7}$--$10^{-6}$. In the SINIS structure, the 
elastic higher-order processes are negligible.

The simplest inelastic process, cotunneling, is energetically forbidden by the BCS gap in the voltage range
$|eV|<2\Delta$. The next-order processes are the Andreev reflection and Cooper-pair/electron cotunneling. The 
Andreev reflection causes a higher error rate, but in the case $E_\mathrm{C}/\Delta=\epsilon_\mathrm{C}>1$, an 
optimum pumping sequence exists, where the Andreev reflection vanishes. In the optimum sequence, the bias 
voltage is $eV/\Delta=1$, and the gate voltage is a square wave with offset $n_\mathrm{g0}=1/2$ and amplitude
$A_\mathrm{g}$ between $1/4\epsilon_\mathrm{C}<A_\mathrm{g}<1/2-1/4\epsilon_\mathrm{C}$. The amplitude should
be closer to the upper end of this range to increase the tunneling rates. Finally, the leakage is
dominated by Cooper-pair/electron cotunneling.

\begin{figure}
\resizebox{0.6\columnwidth}{!}{
  \includegraphics{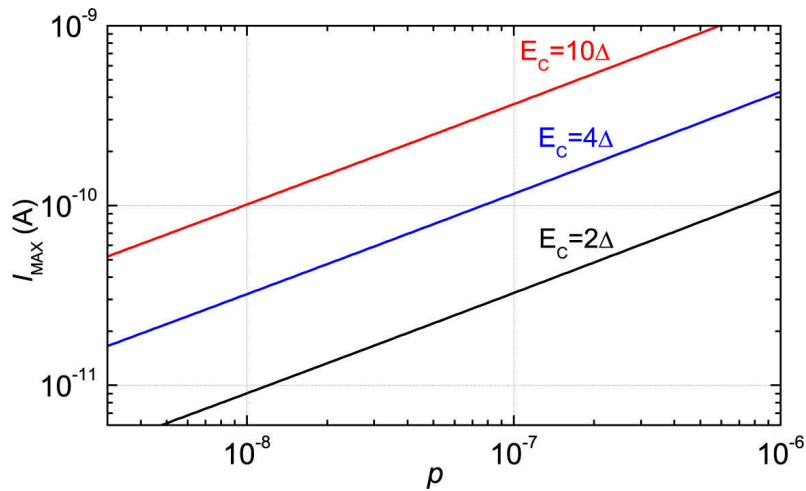} }
\caption{Maximum turnstile current as a function of the allowed error rate $p$ for different sample parameters.}
\label{fig:teorvirhe}
\end{figure}

Reference~\cite{Averin2008} presents estimates for the obtainable error rates of the turnstile with different
sample parameters. They are reproduced in Fig.~\ref{fig:teorvirhe}. The error decreases significantly, when the 
charging energy is increased over the BCS gap energy. At $E_\mathrm{C}/\Delta=4$ and with junction resistance 
400~k$\mathrm{\Omega}$, 30~pA can be reached with the error rate $10^{-8}$. If $E_\mathrm{C}/\Delta=10$
and resistance is 260~k$\mathrm{\Omega}$, 100~pA can be reached with the same accuracy. 
These are still realistic parameters since they were reached, e.g., in Ref.~\cite{Pashkin2000}.

The optimum resistance of the sample is a tradeoff between low leakage (high resistance) and low $RC$ time 
constant. The absence of inelastic cotunneling allows to use lower resistance than in normal state devices. 
Hence, the maximum current can be slightly higher than in normal-state pumps. Also, the capacitance of the
island can be very small, since coupling the gate voltage to the island does not create any cross coupling
to other islands as in multi-island devices.

The error rates presented in Ref.~\cite{Averin2008} do not include the Cooper-pair breaking
rate $\gamma$ of Eq.~(\ref{nS}). However, there seems to be no experimental evidence of $\gamma$ in the case of
high-quality aluminum; the leakages observed in experiments can be caused by Andreev reflection, which also 
causes a linear leakage.
In addition, nonidealities in tunnel barriers like pinholes can cause a linear leakage. In this paper, we use a 
general leakage parameter $\eta=R_\mathrm{T}/R_0$ to avoid confusion with $\gamma$. Note the factor of two 
difference in the Coulomb-blockaded case between the leakage by $\eta$ and that of $\gamma$ presented in
Sec.~\ref{sec:theory}.

Let us consider also the change of the temperature of the
normal-metal island during pumping. The superconducting leads are well thermalized to the base temperature of the
cryostat, but the temperature of the electrons on the island can vary significantly. The heat fluxes arising
from tunneling are proportional to the energy deposition and extraction rates of incoming and outgoing
electrons:
\begin{equation}\label{heatrates} \begin{array}{l}
\dot{Q}^{i,+}=\frac{\Delta^2}{e^2R_{\mathrm{T},i}}\int d\epsilon (\epsilon-\epsilon^{i,+}) n_\mathrm{S}(\epsilon)f_\mathrm{S}(\epsilon)[1-f_\mathrm{N}(\epsilon-\epsilon^{i,+})]\\
\dot{Q}^{i,-}=\frac{\Delta^2}{e^2R_{\mathrm{T},i}}\int d\epsilon (\epsilon+\epsilon^{i,-}) n_\mathrm{S}(\epsilon)f_\mathrm{N}(\epsilon+\epsilon^{i,-})[1-f_\mathrm{S}(\epsilon)]. \end{array}
\end{equation}
The island can cool down at the first two current plateaus in a frequency range limited by the $RC$ time 
constant~\cite{Pekola2007}. At higher frequencies, the island is heated, which can increase the error
rates. In the case $E_\mathrm{C}/\Delta=10$ and $R_\mathrm{T}=260$~k$\mathrm{\Omega}$, cooling is possible up 
to about 10~GHz, i.e., at all envisioned pumping frequencies. The finite temperature of the island causes a 
maximum leakage of $\eta(T)=\sqrt{\pi\Delta/2k_\mathrm{B}T}e^{-\Delta/k_\mathrm{B}T}$ at $n_\mathrm{g}=1/2$, 
which is, however, negligible at about 100~mK.

It should also be noted that at higher current steps, the system should have enough time to set to each charge 
state between the extremes. A staircase-like waveform that waits at each integer gate charge would be equally 
good as the square wave that can be used at the first current plateau. However, the staircase-like waveform is 
not experimentally an obvious choice. Thus, the first step is the most promising for metrology in this respect.

\section{Experimental results}
\label{sec:experiments}

\begin{figure}
\resizebox{0.4\columnwidth}{!}{
  \includegraphics{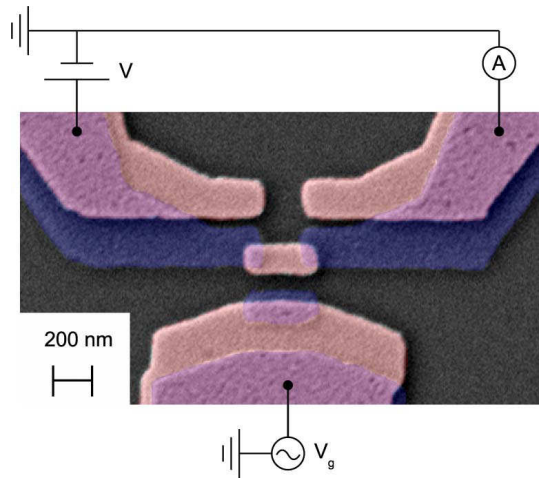} }
\caption{Scanning electron micrograph of the SINIS turnstile. The barely visible grainy
metal is aluminum. The smooth bright metal is copper. The grainy bright surface corresponds to copper 
evaporated on aluminum.}
\label{fig:sample}
\end{figure}

The sample presented in Fig.~\ref{fig:sample} was fabricated by electron beam lithography and two-angle 
evaporation on oxidized silicon wafer. We use the standard PMMA and copolymer as resists. First, we evaporate 
the aluminum superconducting leads. Next, we let oxygen in the vacuum chamber of the evaporator to 
oxidize the surface of the aluminum. The aluminum oxide acts as the tunnel barrier. Finally, we evaporate copper
in a different angle, and we get the same image of the mask as in the aluminum case, but shifted
on the axis of the angle change. Tunnel junctions with lateral size below $100\times 100\; \mathrm{nm}^2$ take 
form where the copper island touches the oxidized aluminum leads. Due to the two-angle technique, we also get 
copper leads and an aluminum island, but the first ones are not connected, and the latter one touches the copper 
end of the gate line and thus becomes a part of the gate capacitor lead.

The resistance of the sample was 475~k$\mathrm{\Omega}$ and the gate capacitance was 14~aF. The 
charging energy was 2.2~K, which is of the order of the BCS gap. Hence, the Andreev reflection was still 
present, and metrological accuracy would not be possible with the present device. In the future, our 
process must be improved to be able to fabricate smaller junctions with high yield. One possibility is to use a 
germanium mask between PMMA and the copolymer~\cite{Pashkin2000}.

The sample was cooled down in a dilution refrigerator to a base temperature of about 60~mK. We used
Thermocoax~\cite{Zorin1995}
as the dc lines between the sample stage and the 1.5~K plate of the refrigerator. This is to avoid sub-gap
leakage caused by photon-assisted tunneling arising from the high-frequency thermal noise that always exists in 
the higher temperature parts of the measurement circuit. Furthermore, we added 550~$\mathrm{\Omega}$ resistors 
on the sample stage on each side of the sample to reduce noise at lower frequencies. The gate couples to the 
sample only via weak capacitances. Hence, the noise properties of the gate line are not as critical as those of 
the dc lines.

\begin{figure}
\resizebox{0.7\columnwidth}{!}{
  \includegraphics{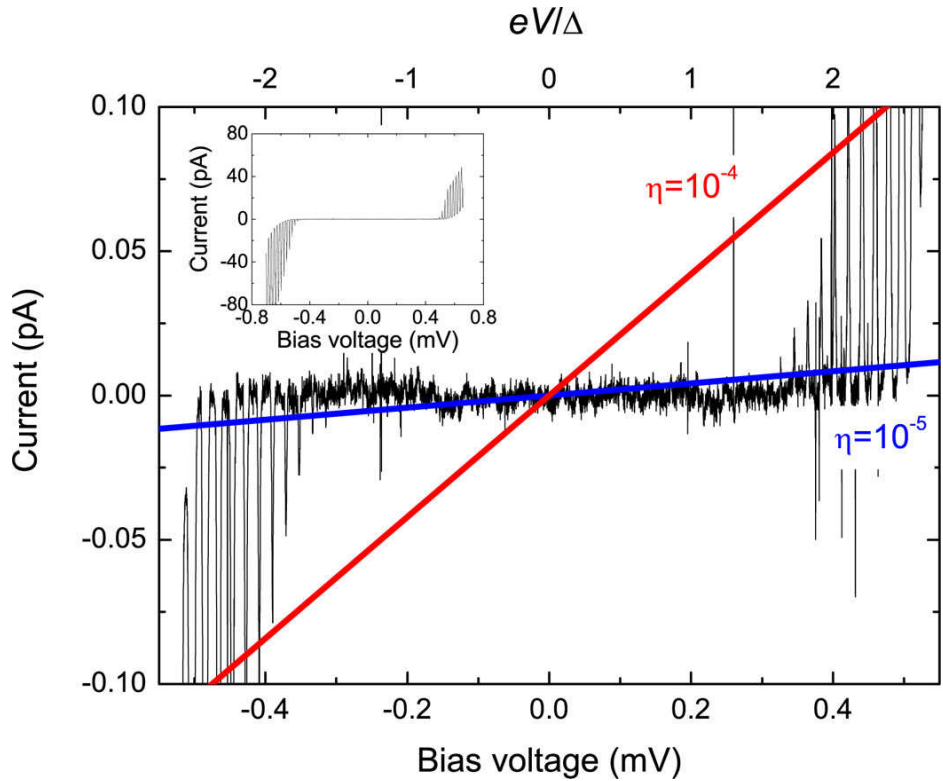} }
\caption{Magnified measured dc IV curve of the sample with resistance 475~k$\mathrm{\Omega}$ and
$E_\mathrm{C}/k_\mathrm{B}$=2.2~K. The inset shows the IV curve of the sample on large scale. The 
gate voltage was swept with constant speed during the IV measurement. Hence, at high voltages, the IV 
curve shows zigzag between the gate open and the gate closed states. At voltages $|eV|<2\Delta$, the gate
periodicity disappears due to the BCS gap. No sub-gap leakage can be seen at a noise level of about
10~fA. The lines correspond to sub-gap leakages $\eta=10^{-4}$ and $10^{-5}$.}
\label{fig:ivyloszoom}
\end{figure}

The measured dc IV curve of the sample is presented in Fig.~\ref{fig:ivyloszoom}. The $\eta$ 
parameter of the sample is clearly below $10^{-5}$, which is a significant improvement to the previous 
measurements with $\eta=1.3\times 10^{-4}$~\cite{Pekola2008}. In the SINIS coolers with large junctions, the 
leakage is typically between $10^{-4}$ and $10^{-3}$ even with highly filtered dc lines similar to ours.
The Coulomb blockade suppresses the leakage caused by the Cooper-pair breaking only by a factor of two. Hence,
our measurement supports the judgement of the theoretical work~\cite{Averin2008} that the leakage is usually 
caused by the Andreev reflection. Large junctions might also have pinholes that are statistically less probable
in our case with small junctions.

The measurement presented in Fig.~\ref{fig:ivyloszoom} shows clearly that verifying the extremely low 
leakages, promised by the theory, will be an experimental challenge. Already at the present stage we are limited 
by the noise of the current measurement.

\begin{figure}
\resizebox{1.0\columnwidth}{!}{
  \includegraphics{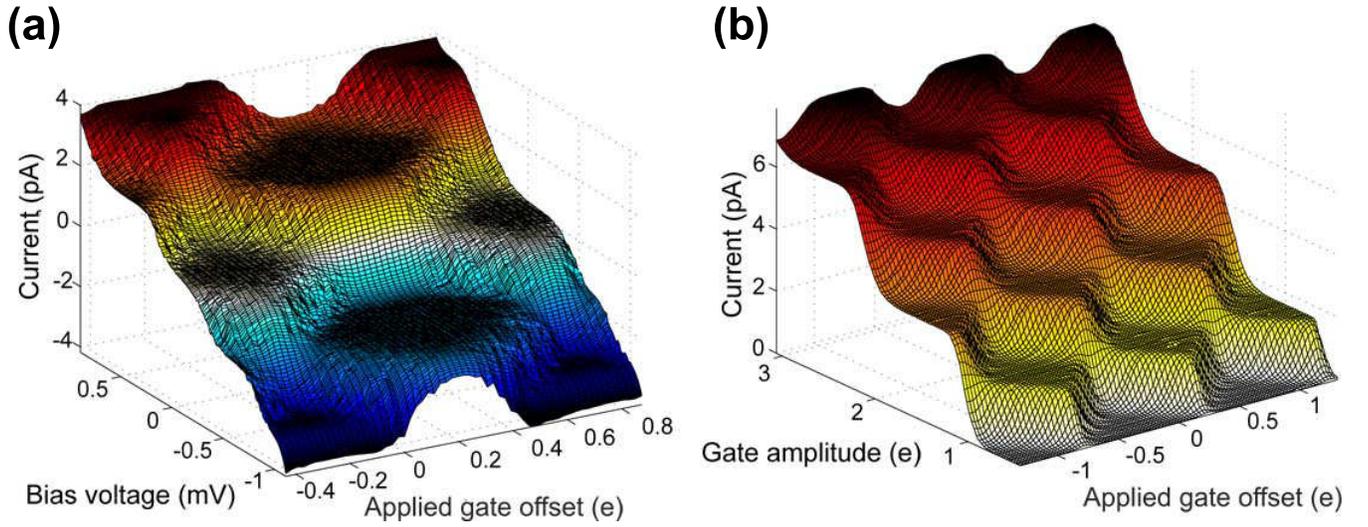} }
\caption{Current plateaus at $f=$10~MHz as a function of (a) bias voltage $V$ and applied gate offset, and (b) 
gate amplitude $A_\mathrm{g}$ and applied gate offset. In (a), $A_\mathrm{g}\approx0.5$, which is suitable for 
pumping one electron in a cycle. When the total offset charge due to the applied offset and the background is 
about 0.5, the gate varies between the charge states 0 and -1. Then, at voltages $eV/\Delta\approx 1$ and
$eV/\Delta\approx -1$, we observe the current plateaus $I=ef\approx$1.6~pA, and $I\approx -1.6$~pA, 
respectively. When the total offset is an integer number, the gate amplitude is not large enough to change the 
charge state, and we observe a zero current plateau. In (b), the sample is biased to $eV/\Delta=1$. When the 
total offset is a half integer, we obtain the odd current plateaus $I=ef$, $I=3ef$ etc.~by increasing the gate 
amplitude. The even steps $I=0$, $I=2ef$ etc.~occur at integer offsets.}
\label{fig:3dkuvat}
\end{figure}

Pumping results of the SINIS turnstile at 10~MHz are presented in Fig.~\ref{fig:3dkuvat}. We observe 
broad quantized current plateaus as in Ref.~\cite{Pekola2008}, but this time with a SINIS structure. However,
the high-frequency gate line did not function perfectly in the present measurement. Above 20~MHz, 
the gate signal was significantly weakened by the line. Even at lower frequencies, the gate signal appeared to 
heat the sample, which rounds the edges of the current plateaus to some extent. Since the square-wave signal 
suggested in Ref.~\cite{Averin2008} requires a broad bandwidth for the higher harmonics of the pumping frequency,
we did not try to employ it here. A sinusoidal signal
$n_\mathrm{g}(t)=n_\mathrm{g0}+A_\mathrm{g}\sin (2\pi ft)$ was used instead. The improvement in the accuracy by 
optimizing the shape of the signal would be very small at the present stage of the development of the turnstile.

\begin{figure}
\resizebox{0.7\columnwidth}{!}{
  \includegraphics{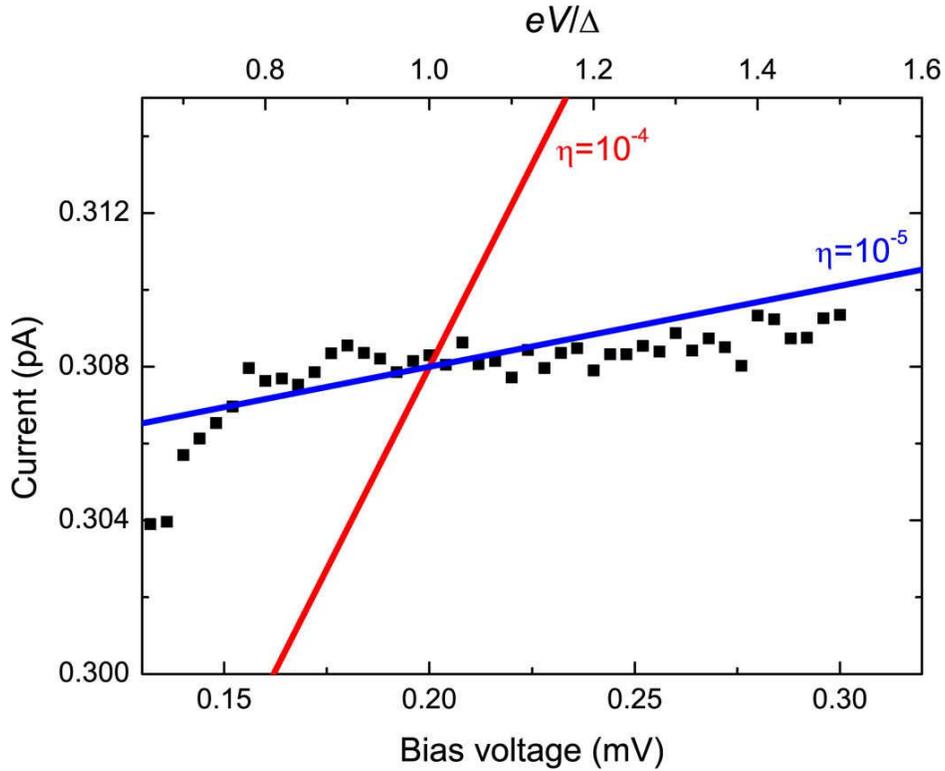} }
\caption{First quantized current step as a function of the bias voltage at 2~MHz pumping 
frequency. Slopes corresponding to sub-gap leakages $\eta=10^{-4}$ and $10^{-5}$ are presented with the 
lines.}
\label{fig:pumppuporras}
\end{figure}

The first step at 2~MHz as a function of the bias voltage is presented in Fig.~\ref{fig:pumppuporras}.
The result is an average of 30 IV measurements. The data points of each IV measurement were measured in 
randomized order to prevent a slope due to a possible drift of the current amplifier. The whole measurement 
lasted for almost two hours. There is no observable slope in the range 0.9--1.2 $eV/\Delta$:
$\eta=(-1.2\pm 3.4)\times 10^{-6}$ ($k=2$, 95\% confidence). This supports the deduction that the
slope of the step 
observed in Ref.~\cite{Pekola2008} was due to the sub-gap leakage. In Ref.~\cite{Pekola2008}, the slope was almost constant over a broad bias range $0.5<eV/\Delta <1.5$. Here, the heating effect rounds the edges
and thus narrows the step. At higher frequencies, there is a slope even at the optimum bias $eV/\Delta=1$,
presumably due to heating. At 10 MHz, the slope corresponds to $\eta\approx 1.3\times 10^{-4}$, which
is about the same as in Ref.~\cite{Pekola2008}.

The theoretical value of the current is about 0.32~pA, but the absolute accuracy of the current amplifier is 
very poor at this level. A traceable measurement with higher accuracy is planned could be performed e.g.~by 
using an electrometer with an external feedback by a standard air capacitor~\cite{Fletcher2007}.

\section{Conclusions}
\label{sec:conclusions}

We have demonstrated the current quantization of the SINIS turnstile. We observe no leakage within
the BCS gap, which supports the optimism of the theoretical error calculations of Ref.~\cite{Averin2008}. The 
upper limit of the ratio between the asymptotic resistance and the sub-gap resistance is clearly below
$10^{-5}$, which is a significant improvement to the
earlier measurements of the turnstile in Ref.~\cite{Pekola2008}.

Theoretically, 100~pA current with a relative accuracy $10^{-8}$ appears to be possible even with a
single turnstile with only one island. The fabrication process must be improved to be able to make
samples with higher charging energies and more transparent junctions. The demands are realistic,
since such sample parameters have been achieved by other research groups~\cite{Kemppinen2009}.

The theoretical potential of the SINIS turnstile has already been studied quite thoroughly. However,
lots of experimental work is still to be done, e.g., a more conclusive experimental study of the leakage 
processes is necessary. The present pumping accuracy can be improved quite easily in an improved set-up as 
regards to the gate rf line. The progress from the idea of a hybrid turnstile to considerable experimental 
results has been fast, though, and the SINIS turnstile is a promising candidate for the quantum current standard.

We acknowledge the Finnish Academy of Science and Letters, Vilho, Yrj\"{o} and Kalle V\"{a}is\"{a}l\"{a}
Foundation, Technology Industries of Finland Centennial Foundation, and the Academy of Finland for financial support.


\end{document}